\documentclass[lettersize,journal]{IEEEtran}
\usepackage{amsmath,amsfonts}
\usepackage{algorithmic}
\usepackage{algorithm}
\usepackage{amsmath}
\usepackage{amssymb}
\usepackage{array}
\usepackage[caption=false,font=normalsize,labelfont=sf,textfont=sf]{subfig}
\usepackage{textcomp}
\usepackage{stfloats}
\usepackage{url}
\usepackage{verbatim}
\usepackage{graphicx}
\usepackage{cite}
\usepackage{mathrsfs}
\usepackage{svg}
\usepackage{upgreek}
\usepackage{balance}
\usepackage{color}
\begin{document}

\title{Average Achievable Rate Analysis for Cell-Free Massive MIMO in the Finite Blocklength Regime with Imperfect CSI}

\author{~Kai~Chen,~\IEEEmembership{Student~Member,~IEEE,}~Feng~Ye,~\IEEEmembership{Student~Member,~IEEE,}~Jiamin~Li,~\IEEEmembership{Member,~IEEE,}
~Dongming~Wang,~\IEEEmembership{Member,~IEEE,}~Pengcheng~Zhu,~\IEEEmembership{Member,~IEEE,}~Xiaohu~You,~\IEEEmembership{Fellow,~IEEE,}
\thanks{This work was supported in part by the National Key R\&D Program of China under Grant 2021YFB2900300, by the Fundamental Research Funds for the Central Universities under Grant 2242022k60006, by the Major Key Project of PCL (PCL2021A01-2), and by the Postgraduate Research\&Practice Innovation Program of Jiangsu Province under Grant KYCX24\_0408. \emph{(Corresponding author: Jiamin Li.)}}
\thanks{The authors are with the National Mobile Communications Research Laboratory, Southeast University, Nanjing {\rm 210096}, China (e-mail: \{220230969; yefeng; jiaminli; wangdm; p.zhu; xhyu\}@seu.edu.cn). J. Li, D. Wang, P. Zhu and X. You are also with Purple Mountain Laboratories, Nanjing, {\rm 211111}, China.}}
\markboth{Journal of \LaTeX\ Class Files,~Vol.~XX, No.~X, XX~XXX}%
{Shell \MakeLowercase{\textit{et al.}}: A Sample Article Using IEEEtran.cls for IEEE Journals}



\maketitle

\begin{abstract}
Acquiring perfect channel state information (CSI) is challenging due to the increasing number of channel parameters in CF-mMIMO systems, especially in uRLLC scenarios. Moreover, the impact of imperfect CSI on the average achievable rate, particularly in the finite blocklength regime, remains underexplored. To this end, this paper proposes a novel approach that derives an accurate closed-form expression for the average achievable rate with imperfect CSI in the Laplace domain. We demonstrate that the expectation and variance of both channel dispersion and channel capacity can be reformulated using the Laplace transform of the large-scale fading term. The simulation results illustrate that closed-form expressions derived in this paper align closely with the Monte Carlo results. Subsequently, the impact of the imperfect CSI on the system performance of CF-mMIMO in the finite blocklength regime is theoretically analyzed. It shows that the imperfect CSI significantly affects the average achievable rate in the finite blocklength regime, while the CF-mMIMO architecture can effectively compensate for the performance loss caused by imperfect CSI. 
\end{abstract}


\section{Introduction}
\IEEEPARstart{T}{he} next-generation applications are expected to encompass a wide range of advancements, including smart technologies and immersive extended reality (XR), aiming to meet the evolving needs of individual users and various industries. These advancements drive an increasing demand for ultra-reliable low latency communication (uRLLC), and lead to a significant increase in key performance indicators (KPIs) of 6G.

Cell-free massive MIMO (CF-mMIMO) has emerged as a promising sixth-generation (6G) technology \cite{14} to meet the future ubiquitous connectivity requirements and ever-growing data traffic demands. In CF-mMIMO system, a large number of distributed access points (APs) jointly serve all the UEs by coherent joint transmission and reception \cite{15} by leveraging a central processing unit (CPU) through fronthual connections. CF-mMIMO incorporates the benefits of massive MIMO (mMIMO) and a distributed system. On the one hand, mMIMO technology improves network capacity by increasing the number of parallel links in the spatial domain \cite{29}. On the other hand, cell-free architecture provides a significantly greater advantage in managing interference by eliminating the boundary effects existing in cellular networks \cite{16}. Moreover, cell-free architecture also has the potential to support uRLLC traffic \cite{18}, because it is beneficial to minimize the average distance between users and APs by deploying more distributed APs in the uRLLC regime. 
	
However, one of the major challenges of uRLLC is instantaneous channel state information (CSI) acquisition. In practical scenarios, perfect CSI is difficult to obtain in CF-mMIMO systems for the uRLLC regime. On the one side, implementing uRLLC relies on short packet lengths and small transmission time intervals, leaving fewer bits for CSI acquisition overhead. On the other side, an increased number of channel parameters in CF-mMIMO systems making it difficult to acquire high-accuracy CSI \cite{12}. Since the imperfect CSI is inevitable in practical scenarios, several attempts to account for imperfect CSI on system performace can be found in the literature.

Imperfect CSI has been well-studied in \cite{8}, \cite{1}, \cite{22} and \cite{17}. Specifically, 
\cite{8} studied both lower and upper bounds of mutual information with imperfect CSI in a MIMO system for Gaussian inputs. Similarly, \cite{1} demonstrated that the mutual information increases with the number of antennas, but is limited by imperfect CSI in high SNR. Since CSI plays an important role in power allocation, \cite{22} investigated the optimal transmitter strategies with imperfect CSI in ergodic MIMO channels. The above studies primarily focus on the impact of imperfect CSI in MIMO systems. Meanwhile, for a CF-mMIMO system, \cite{17} provided a comprehensive analysis of CF-mMIMO with imperfect CSI by comparing four different implementations of CF-mMIMO and showing that a centralized implementation with optimal MMSE processing could maximize the SE. 

Although the aforementioned studies provide comprehensive analyses of the impact of imperfect CSI on system performance, they fall short in providing a precise closed-form approximation. This limitation makes it challenging to analyze the relationship between system performance and system parameters effectively. Notably, the closed-form expression of the achievable rate for an mMIMO system with imperfect CSI was first derived by \cite{23} for both maximum-ratio combining and zero-forcing receivers. Similarly, \cite{24} provided the closed-form expressions for spectral efficiency under imperfect CSI in the uplink CF-mMIMO system.

However, the above analyses predominantly focus on imperfect CSI in general scenarios without considering uRLLC requirements and the impact of imperfect CSI on the average achievable rate. To accurately design and evaluate system parameters, it is crucial to develop concise closed-form approximations that directly reflect the relationship between imperfect CSI and average achievable rate in a CF-mMIMO system with finite blocklength. In the finite blocklength regime, traditional information-theoretic metrics, such as Shannon capacity, are unsuitable \cite{19}. Polyanskiy et al. investigated a unified approach to obtain tight bounds on the maximal achievable rate using normal approximation \cite{20}. Considering imperfect CSI, \cite{8} investigated the achievable rate with imperfect CSI under both ergodic and non-ergodic scenarios in a MIMO system. These studies lay a foundation for evaluating the impact of imperfect CSI on average achievable rate in CF-mMIMO systems within the finite blocklength regime.
 
Based on the above analysis, the main objectives of this paper are to derive explicit closed-form expressions for the average achievable rate in the CF-mMIMO system with imperfect CSI for finite blocklength, analyze the relationship between system performance and imperfect CSI, and theoretically explore ways to compensate for the system performance reduction caused by imperfect CSI in the CF-mMIMO system. The main contributions of this paper are as follows:
\begin{enumerate}
\item We proposes a novel approach that derives an accurate closed-form expression for the average achievable rate with imperfect CSI by Laplace approximation. We prove that the expectation and variance of both channel dispersion and channel capacity can be reformulated as integrals involving the Laplace transform of the large-scale fading term. And then the Laplace transform of the large-scale fading term is derived.
\item The impact of imperfect CSI on the performance of the CF-mMIMO system under finite blocklength is theoretically analyzed in this paper. We demonstrate that imperfect CSI significantly affects both the average achievable rate and channel stability within the CF-mMIMO system. Furthermore, the growth of the normalized achievable rate with respect to SNR becomes limited when CSI is imperfect.
\item The advantage of CF-mMIMO over centralized MIMO in terms of system performance with respect to imperfect CSI is demonstrated in this paper. It is clearly proved that mMIMO technology can compensate for the reduction in the achievable rate caused by imperfect CSI by increasing the number of spatial streams. Moreover, the cell-free architecture ensures that channels with various channel estimation errors converge towards the same stability as deterministic channels.
\end{enumerate}

The rest of this paper is organized as follows. Section \ref{sec:1} defines the system model for uplink CF-mMIMO with imperfect CSI in finite blocklength regime. Next, Sections \ref{sec:2} analyze the statistical properties of channel capacity and channel dispersion with imperfect CSI. Section \ref{sec:4} analyzes the system performance under various channel estimation error levels and presents a clear comparison between perfect CSI and imperfect CSI. Finally, the major conclusions are drawn in Section \ref{sec:5}.

\textit{Notation}: Boldface letters are utilized to represent matrices (uppercase) or vectors (lowercase), whereas calligraphy letters are reserved for collections. The mean and variance of a random variable are described using the operators $\mathbb E\left( \cdot \right)$ and ${\text{Var}}\left( \cdot \right)$, respectively. ${\mathbb{C}}^{M \times N}$ is used to denote the set of complex matrices with dimension ${M \times N}$. The notations $\left( \cdot \right)^{\rm T}$ and $\left( \cdot \right)^{\rm H}$ are employed to represent the transpose and the conjugate transpose of a vector or matrix, respectively. Additionally, the trace and determinant of a matrix are denoted by $\text{tr} \left( \cdot \right)$ and $\text{det} \left( \cdot \right)$, respectively. The Frobenius norm of a matrix $\bf X$ is denoted by ${\left\| {\bf{X}} \right\|_{\text{F}}} = \sqrt {{\text{tr}}\left( {{\bf{X}}{{\bf{X}}^{\text{H}}}} \right)} $. An N-dimensional identity matrix is denoted as ${\bf{I}}_N$. The distribution $\mathcal{CN} \left( \mu,\sigma^2 \right)$ is introduced to represent the circularly symmetric complex Gaussian distribution with mean $\mu$ and variance $\sigma^2$. The kronecker product is denoted by $\otimes$. Lastly, ${}_p\psi_q$ denotes Fox-H function.

\section{System Model}
\label{sec:1}
We consider an uplink CF-mMIMO network, comprising $N$ APs distributed across a two-dimensional Euclidean plane \cite{2}. The APs follow a Poisson Point Process (PPP) distribution $\Phi$ with a mean density of $\lambda$ (measured in APs per square meter), and each AP is equipped with $L$ antennas. All APs are connected to a CPU via fronthaul links and collaboratively serve users, where each user is equipped with $M$ antennas. 

We define $\omega = L/M$ as the ratio of the number of AP antennas to the number of user antennas. Since the spatial degree of freedom (DoF) represents the minimum of the number of transmitting and receiving antennas, the number of user antennas $M$ is the spatial DoF \cite{3}.

To simplify the model, we consider a circular region with radius $R$. Thus, the average number of APs is
\begin{equation}
\mathbb{E}\left[{N}\right] = {\lambda \pi R^2}\buildrel \Delta \over={E_N}.
\end{equation}

Reference \cite{6} demonstrates that when solving for the achievable rate of the system by treating it as a large MIMO network, the problem can be reformulated as the sum of the achievable rates of multiple independently coded users. Thus, in this scenario, we analyze the case of a randomly selected single user, since the performance observed by any user remains consistent regardless of their location under a PPP distribution. The performance of the CF-mMIMO system is then evaluated by focusing on the set of APs serving this user as the central point.

\subsection{Receive Signal Model with imperfect CSI}
We consider flat-fading quasi-static channels in the finite-length regime \cite{4}, where the channel coefficients remain constant during a packet transmisson time of $\tau$. It is assumed that the receiver is provided with \textcolor{blue}{imperfect CSI by joint decoding} and the transmitter has no CSI, since the receiver may not have sufficient time to provide feedback on CSI in low-latency scenarios.  Thus, according to \cite{17} and \cite{1}, the channel can be expressed as the sum of its minimum mean squared error (MMSE) estimator, ${\widehat {\bf{h}}_n}$ , and an error term, ${{\bf{E}}_n}$, as
\begin{equation}
{{\bf{h}}_n} = {\widehat {\bf{h}}_n} + {{\bf{E}}_n}.
\end{equation}
Then ${\widehat {\bf{h}}_n} \in \mathbb{C}^{L \times M}$ and ${{\bf{E}}_n} \in \mathbb{C}^{L \times M}$ are uncorrelated due to the property of MMSE estimation. The entries of ${{\bf{E}}_n}$ are i.i.d and follow ${\cal C}{\cal N}\left( {0,\sigma_e^2} \right)$, i.e., $\mathbb{E}\left\{ {{E_{i,j}}{E_{m,n}}} \right\} = \sigma _e^2{\delta _{i - m,j - n}}$. Similarly, the entries of ${{\widehat {\bf{h}}_n}}$ are i.i.d and follow ${\cal C}{\cal N}\left( {0,1-\sigma_e^2} \right)$.

In such a system, the uplink receive signal at the CPU can be expressed as:
\begin{equation}
{\bf{Y}} = \widehat {\bf{H}}\left( {\bf{d}} \right){\bf{X + EX}} + {\bf{W}},
\end{equation}
where ${\bf{Y}} \in {\mathbb{C}^{NL \times \tau }}$ is the discrete-time received signal over $\tau$ samples, ${\bf{X}} \in {\mathbb{C}^{M \times \tau }}$ is the transmitted signal matrix satisfying power constraint $\left\| {\bf{X}} \right\|_{\rm{F}}^2 \le \tau \rho $, and $\rho$ is the signal-to-noise (SNR). Moreover, ${\bf{E}} \in {\mathbb{C}^{NL \times M}} = {[ {\sqrt {{l}\left( {{d_1}} \right)} {\bf{E}}_{\rm{1}}^{\rm{T}}, \ldots ,\sqrt {{l}\left( {{d_n}} \right)} {\bf{E}}_N^{\rm{T}}} ]^ {\rm{T}}}$ and the channel matrix  ${\widehat{\bf{H}}}\left( {\bf{d}} \right) \in {\mathbb{C}^{NL \times M}}= {[ {\sqrt {{l}\left( {{d_1}} \right)} {\widehat{\bf{h}}_1^{\rm T}}, \ldots ,\sqrt {{l}\left( {{d_N}} \right)} {\widehat{\bf{h}}_N^{\rm T}}} ]^{\rm{T}}}$ is parameterized by the distance vector ${\bf{d}} = {\left[ {{d_1}, \ldots ,{d_N}} \right]^{\rm T}}$. The large-scale fading $l\left( {{d_n}} \right)$ is modeled as:
\begin{equation}
{l}\left( {{d_n}} \right) = \min \left\{ {1,{d_n^{ - \alpha }}} \right\},
\end{equation}
where $\alpha=3.7$ is the pass loss exponent \cite{7}. Furthermore, ${\bf{W}} \in \mathbb{C}{^{NL \times \tau }}$ represents the additive white gaussian noise (AWGN). It is assumed that $\bf{W}$ is independent of  $\widehat{\bf{H}}\left(\bf d \right)$ and $\bf E$, with entries i.i.d as ${\cal C}{\cal N}\left( {0,1} \right)$. 

In the absence of CSI at the transmitter \cite{5}, the optimal power allocation should fulfill:
\begin{equation}
\frac{1}{\tau}{\bf{X}}{{\bf{X}}^{\rm{H}}} = {\rho  \over M}{{\bf{I}}_M} \buildrel \Delta \over = {{\bf{\Sigma }}_{\bf{X}}}.
\label{eq:eq2}
\end{equation}
Then we have
\begin{equation}
\begin{aligned}
\frac{1}{\tau} \mathbb{E} \left[ {{\bf{EX}}{{\bf{X}}^{\rm H}}{{\bf{E}}^{\rm H}}} \right] &= \frac{\rho }{M}\mathbb{E}\left[ {{\bf{E}}{{\bf{E}}^{\rm H}}} \right] \\
&= \rho \sigma _e^2 \mathbb{E}\left[ \operatorname{diag} \left( {{l}\left( {{d_1}} \right), \ldots ,{l}\left( {{d_N}} \right)} \right) \otimes {{\bf{I}}_L} \right] \\
&= \rho\sigma _e^2 \mathbb{E} \left[ l \left( d \right) \right] {{\bf{I}}_{NL}}\\ \label{eq:EX}
&= \rho\sigma _e^2 \theta {{\bf{I}}_{NL}} \\
 &\buildrel \Delta \over = \bf \Sigma_{EX},
\end{aligned}
\end{equation}
where, 
\begin{equation}
\begin{aligned}
\mathbb{E} \left[ l \left( d \right) \right] &= \int_0^R {l\left( r \right)\frac{{2r}}{{{R^2}}}} dr =\frac{\alpha-2R^{2-\alpha}}{{\alpha R^2-2R^2}} \triangleq \theta.
\end{aligned}
\end{equation}

\subsection{Normal Approximation}
A lower bound on the achievable rate for imperfect CSI with ergodic channel states is provided in \cite{8} as:
\begin{equation}
{R^*}\left( {\tau ,\varepsilon } \right) \geqslant C({\bf{\widehat{H}}}) - \sqrt {\frac{{V( {\widehat {\bf{H}}} )}}{\tau }} \frac{{{Q^{ - 1}}\left( \varepsilon  \right)}}{{\ln 2}}\label{eq:Rate},
\end{equation}
where, $C(\bf{\widehat{H}})$ is the channel capacity, ${V( {\widehat {\bf{H}}} )}$ is the channel dispersion. $Q\left( \cdot \right)$ represents the Gaussian Q-function and $\varepsilon$ represents the block error probability (BEP). 

To further analyze the achievable rate with imperfect CSI under finite blocklength in CF-mMIMO systems, the statistical properties of the channel capacity $C(\bf{\widehat{H}})$ and the channel dispersion $V(\bf{\widehat{H}})$ with imperfect CSI need to be analyzed separately. 

According to \cite{9}, the lower bound of the channel capacity is given as: 
\begin{equation}
\begin{aligned}
  &C({\bf{\widehat{H}}})= {\log _2}\det \left[ {{{\bf{I}}_M} + \widehat {\bf{H}}{{\left( {\bf{d}} \right)}^{\rm{H}}}{{\left( {{{\bf{I}}_{NL}} + {{\bf{\Sigma_{EX} }}}} \right)}^{ - 1}}\widehat {\bf{H}}\left( {\bf{d}} \right){\bf{\Sigma_X }}} \right] \label{eq:C1}.
\end{aligned}
\end{equation}
By substituting (\ref{eq:EX}) and  (\ref{eq:eq2}) into (\ref{eq:C1}) ,we have
\begin{equation}
\begin{aligned}
 C(\bf{\widehat{H}})&= {\log _2}\det \left[ {{{\bf{I}}_M} + \frac{\rho }{{M + M\rho \sigma _e^2\theta }}\widehat {\bf{H}}{{\left( {\bf{d}} \right)}^{\rm{H}}}\widehat {\bf{H}}\left( {\bf{d}} \right)} \right] \\
  &\buildrel (a) \over= \sum\limits_{i = 1}^M {{{\log }_2}\left( {1 + \frac{{\rho {\lambda _i}}}{{M + M\rho \sigma _e^2\theta }}} \right)} , \label{eq:C}
\end{aligned}
\end{equation}
where $ \widehat {\bf{H}}{{\left( {\bf{d}} \right)}^{\rm{H}}}\widehat {\bf{H}}\left( {\bf{d}} \right) $ is a Hermitian matrix that can be facrotized into ${\bf{Q\Lambda }}{{\bf{Q}}^{\rm{H}}}$, with ${\bf{\Lambda }} = {\text{diag}}\{ {\lambda _1}, \cdots ,{\lambda _M}\} $ being a diagonal matrix containing the eigenvalues of $ \widehat {\bf{H}}{{\left( {\bf{d}} \right)}^{\rm{H}}}\widehat {\bf{H}}\left( {\bf{d}} \right) $ and ${\bf{Q}}\in{\mathbb C}^{M \times M}$ being a unitary matrix. Step $\left(a\right)$ follows from this decomposition and futher employs Sylvester's determinant identity.

According to \cite{4}, the expression for channel dispersion is derived as:
\begin{equation}
V({\bf{\widehat{H}}}) =  M- \sum\limits_{i = 1}^M {{{{{\left( {1 + \frac{{\rho {\lambda _i}}}{{M + M\rho \sigma _e^2\theta }}} \right)}^{-2}}}}} \label{eq:V1}.
\end{equation}

It can be observed from (\ref{eq:C}) that the channel capacity with imperfect CSI can be regarded as a low-SNR scenario. Most previous derivations of the average achievable rate are based on high-SNR assumptions \cite{6}, \cite{3}. However, in imperfect CSI scenarios, the channel capacity and channel dispersion are highly sensitive to approximations, making these derivation methods often inapplicable. To address this, we propose a novel derivation approach for the average achievable rate. In the following sections, the Laplace approximation will be applied to derive precise expressions for both the channel capacity and the channel dispersion.

\section{Statistal Properties of Channel Dispersion and Channel Capacity}\label{sec:2}
In this section, we first demonstrate that the expectation and variance of both channel dispersion and channel capacity can be reformulated as the integrals of the Laplace transform of the large-scale fading term. Subsequently, a theorem providing the closed-form expressions for the expectation and variance of channel dispersion and channel capacity are presented, respectively.

\subsection{Statistal Properties of Channel Dispersion}
Taking the expectation of (\ref{eq:V1}), we have:
\begin{equation}
\begin{aligned}
  {\mathbb E}\left[ {V( {\widehat {\bf H}} )} \right]&= {\mathbb E}\left[ {M - \sum\limits_{i = 1}^M {{{\left( {1 + \frac{{\rho {\lambda _i}}}{{M + M\rho \sigma _e^2\theta }}} \right)}^{ - 2}}} } \right] \\
   &= M - {\mathbb E}\left[ {\sum\limits_{i = 1}^M {{{\left( {1 + \frac{{\rho {\lambda _i}}}{{M + M\rho \sigma _e^2\theta }}} \right)}^{ - 2}}} } \right] \label{eq:V} ,
\end{aligned}
\end{equation}
where, $G \triangleq \sum\limits_{i = 1}^M {{{\left( {1 + \frac{{\rho {\lambda _i}}}{{M + M\rho \sigma _e^2\theta }}} \right)}^{ - 2}}} $ for simplicity. Therefore, we need to solve the expectations on the right-hand-side (RHS), which can be rewritten as
\begin{equation}
\begin{aligned}
 G  ={{\text{tr}}\left[ {{{\left( {{{\bf{I}}_M} + \frac{{\rho }}{{M + M\rho \sigma _e^2\theta }}\sum\limits_{n \in {\Phi}} {l\left( {{d_n}} \right)\widehat {\bf{h}}_n^{\rm H}{{\widehat {\bf{h}}}_n}} } \right)}^{ - 2}}} \right]}. \label{eq:G}
\end{aligned}
\end{equation}

The expectation of (\ref{eq:G}) can be decomposed into independent expectations over the small-scale and large-scale components. Since ${\widehat {\bf{h}}_n^{\rm H}{{\widehat {\bf{h}}}_n}}$ is a central Wishart distribution, the ergodic small-scale information can be acquired using the expectation property of the Wishart distribution:
\begin{equation}
\begin{aligned}
&{\mathbb E_{\bf{h}}}\left[ G \right] \\
&={\mathbb E_{\bf{h}}}\left\{ {{\text{tr}}\left[ {{{\left( {{{\bf{I}}_M} + \frac{{\rho}}{{M + M\rho \sigma _e^2\theta }}\sum\limits_{n \in {\Phi}} {l\left( {{d_n}} \right)\widehat {\bf{h}}_n^{\rm H}{{\widehat {\bf{h}}}_n}} } \right)}^{ - 2}}} \right]} \right\} \\
&{\buildrel \left(a\right) \over \approx} {\text{tr}}\left[ {{{\left( {{{\bf{I}}_M} + \frac{{\rho }}{{M + M\rho \sigma _e^2\theta }}\sum\limits_{n \in {\Phi }} {l\left( {{d_n}} \right){{\mathbb E}_{\bf{h}}}\left( {\widehat {\bf{h}}_n^{\rm H}{{\widehat {\bf{h}}}_n}} \right)} } \right)}^{ - 2}}} \right] \\
   &\buildrel \left(b\right) \over = M{\left[ {1 + \frac{L}{M}\frac{{\rho \left( {1 - \sigma _e^2} \right)}}{{1 + \rho \sigma _e^2\theta }}\sum\limits_{n \in {\Phi }} {l\left( {{d_n}} \right)} } \right]^{ - 2}}  \\
 &= M{\beta ^2}{\left[ {\beta  + \sum\limits_{n \in {\Phi}} {l\left( {{d_n}} \right)} } \right]^{ - 2}}, \label{eq:eq001}
\end{aligned}
\end{equation}
where, step $\left(a\right)$ holds due to the Cauchy-Schwarz inequality \cite{26} while step $\left(b\right)$ follows from Lemma 2.
 $ { {\frac{{1 + \rho \sigma _e^2\theta }}{{\rho\omega \left( {1 - \sigma _e^2} \right)}}} } \triangleq \beta$ for simplicity.

Since the spatial randomness of APs can be perfectly described in the Laplace domain, a high-accuracy expression for the expectation of channel dispersion can be obtained by applying the Laplace approximation. According to Lemma 1, we have 
\begin{equation}
\begin{aligned}
{\mathbb E}\left[ G \right] &= M{\beta ^2} {\mathbb{E_{\bf d}}} \left[ {\left( {\beta  + \sum\limits_{n \in {\Phi}} {l\left( {{d_n}} \right)}} \right)^{ - 2}} \right] \\
&\buildrel (a)\over = M{\beta ^2}{\mathbb E}\left[ {\int_0^\infty  {s{e^{ - s\left( {\beta  + X} \right)}}} ds} \right] \\ 
&\buildrel (b)\over \approx M{\beta ^2}\int_0^\infty  {s{e^{ - s\beta }}{\mathbb E}\left[ {{e^{ - sX}}} \right]} ds \\
&= M{\beta ^2} \int_0^\infty  {s{e^{ - s\beta }}{{\mathcal{L}}_X}\left( s \right)} ds,  \label{eq:eqV2}
\end{aligned}
\end{equation}
where, $\sum_{n \in \Phi } {l\left( {{d_n}} \right)} \triangleq X$. Step $\left(a\right)$ utilizes the Laplace approximation, while step $\left(b\right)$ applies Fubini’s theorem \cite{25} to exchange the order of expectation and integration. ${\mathcal{L}_X\left( s \right)} $ denotes the Laplace transform of $X$.

This demonstrates that calculating the expectation of channel dispersion can be transformed into solving the Laplace transform of $X$, as presented in the following theorem.

\textit{Theorem 1}: For a homogeneous PPP distribution $\Phi$ with density $\lambda$ in a circular region of radius $R$, we can derive the Laplace transform of $X = \sum_{n \in \Phi } {l\left( {{d_n}} \right)}$

\begin{equation}
{{\mathcal{L}}_X}\left( s \right)= \exp \left\{ { - \lambda \pi \left[ {  {{R^2} - {e^{ - s}}} - 2 \int_1^R {{e^{ - s{r^{ - \alpha }}}}rdr} } \right]} \right\}\label{eq:eqqV4}.
\end{equation}
A simplified version is given by:
\begin{equation}
{{\mathcal{L}}_X^{approx.}}\left( s \right) = \exp \left\{ {\frac{{2\pi \lambda }}{\alpha }\Gamma \left( { - \frac{2}{\alpha }} \right){s^{\frac{2}{\alpha }}} + \frac{{2\pi \lambda {R^{2 - \alpha }}}}{{\alpha  - 2}}s} \right\}  \label{eq:eqV4}.
\end{equation}

\textit{Proof}: See Appendix B. 

Note that when using (\ref{eq:eqV4}), the constraint $\beta/\lambda > \left({2 \pi  R^{2 - \alpha}}\right)/\left({\alpha - 2}\right)$ must be satisfied to ensure the convergence of the integral in (\ref{eq:eqV2}). It is obvious that extreme assumptions, such as $N\rightarrow \infty$, typically do not satisfy this constraint. Furthermore, larger channel estimation errors lead to a higher value of $\beta$.

Theorem 1 provides the Laplace transform of large-scale fading term $\sum_{n \in \Phi } {l\left( {{d_n}} \right)}$, which serves as the foundation for deriving the expectation and variance of channel dispersion and channel capacity in this paper. Then, the expectation and variance of channel dispersion can be obtained, as presented in the following theorem.

\textit{Theorem 2}: Accurate closed-form expressions for the expectation and variance of channel dispersion based on Laplace approximation.

The expectation of channel dispersion is expressed as:
\begin{equation}
\begin{aligned}
 {\mathbb E}\left[{V({\widehat {\bf H}} ) }\right] = M - M \beta^2 \int_0^\infty s e^{-s\beta}\mathcal{L}_X\left(s\right)ds \label{eq:VN1},
\end{aligned}
\end{equation}

The variance of channel dispersion is expressed as:
\begin{equation}
\begin{aligned}
{\text{Var}}\left[ {V( {\widehat {\bf{H}}} )} \right] = &\frac{{{M^2}{\beta ^4}}}{6}\int_0^\infty  {{s^3}{e^{ - s\beta }}{{\mathcal{L}}_X}\left( s \right)} ds\\
&- {M^2}{\beta ^4}{\left[ {\int_0^\infty  {s{e^{ - s\beta }}{{\mathcal{L}}_X}\left( s \right)} ds} \right]^2}. \label{eq:eqVVVV}
\end{aligned}
\end{equation}

By applying the simplified Laplace transform (\ref{eq:eqV4}), the expectation of channel dispersion in (\ref{eq:VN1}) can be simplified as
\begin{equation}
\begin{aligned}
 {\mathbb E}\left[{V({\widehat {\bf H}}) }\right] &= M - \frac{M \beta^2}{\left( \beta - \frac{2 \pi \lambda R^{2 - \alpha}}{\alpha - 2} \right)^2} \\
&\quad \times {}_1 \psi_0 \left[ \left( 2, \frac{2}{\alpha} \right); \frac{\frac{2 \pi \lambda}{\alpha} \Gamma\left( - \frac{2}{\alpha} \right)}{\left( \beta - \frac{2 \pi \lambda R^{2 - \alpha}}{\alpha - 2} \right)^{\frac{2}{\alpha}}} \right] \label{eq:VN2}.
\end{aligned}
\end{equation}

Similarly, the variance of channel dispersion can be simplified as shown in (\ref{eq:VarV}) at the top of the next page.
\begin{figure*}[ht]
\centering
\begin{equation}
\begin{aligned}
 &\text{Var}\left[ {V({\widehat{\bf{H}}} ) }\right] =\frac{M^2 \beta^4}{6 \left( \beta - \frac{2\pi \lambda R^{2 - \alpha}}{\alpha - 2} \right)^4} {}_1\psi_0 \left[  \left( 4, \frac{2}{\alpha} \right)  ; \frac{\frac{2\pi \lambda}{\alpha} \Gamma \left( -\frac{2}{\alpha} \right)}{\left( \beta - \frac{2\pi \lambda R^{2 - \alpha}}{\alpha - 2} \right)^{\frac{2}{\alpha}}} \right]- \left\{ \frac{M \beta^2}{\left( \beta - \frac{2\pi \lambda R^{2 - \alpha}}{\alpha - 2} \right)^2} {}_1\psi_0 \left[  \left( 2, \frac{2}{\alpha} \right)  ; \frac{\frac{2\pi \lambda}{\alpha} \Gamma \left( -\frac{2}{\alpha} \right)}{\left( \beta - \frac{2\pi \lambda R^{2 - \alpha}}{\alpha - 2} \right)^{\frac{2}{\alpha}}} \right] \right\}^2 \label{eq:VarV}
\end{aligned}
\end{equation}
\hrulefill
\end{figure*}

\textit{Proof}: See Appendix C.

 \begin{figure}[ht]
 \centering
 \includegraphics[width=3 in]{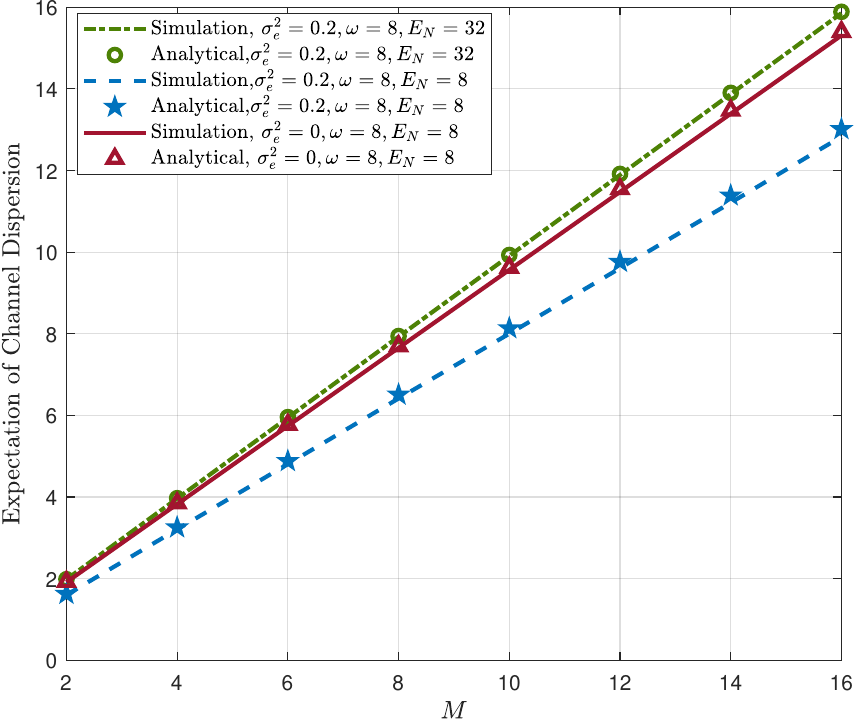}
 \caption{Performace comparison of simulation and analytical values of channel dispersion with $E_N=8, \gamma=-20\ {\rm{dB}}, R=50\ m$.}
 \label{fig1}
 \end{figure}
Fig.\ref{fig1} shows the fitting between the simulation and analytical values for expected channel dispersion, with the analytical values obtained from (\ref{eq:VN1}) according to the constraint mentioned in (\ref{eq:eqV4}). $\gamma \buildrel \Delta \over = \rho r^{-\alpha}$ represents the SNR from the typical user to the region boundary. It illustrates that expected channel dispersion increases with higher spatial DoF. When perfect CSI is available, the expected channel dispersion closely aligns with the DoF. However, when the number of APs is insufficient, imperfect CSI results in a significant reduction in the expected channel dispersion. Then, as the number of APs increases, the channel dispersion under imperfect CSI progressively approaches the ideal scenario observed with perfect CSI. This trend also holds true for larger $\omega$.

 \begin{figure}[ht]
 \centering
 \includegraphics[width=3 in]{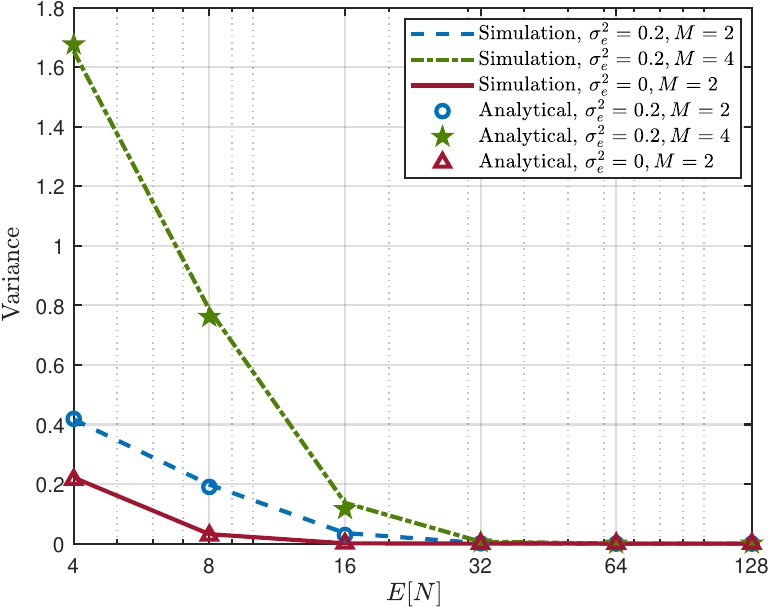}
 \caption{Performace comparison of simulation and analytical values of variance of channel dispersion with $\omega=8, \gamma=-20\ {\rm{dB}}, R=50 m$.}
 \label{fig2}
 \end{figure}
Fig. \ref{fig2} shows the fitting of the simulation and analytical values for the variance of channel dispersion, with the analytical values obtained from (\ref{eq:eqVVVV}). The variance of channel dispersion increases with higher spatial DoF due to the fact that more independent channels are affected by distortion. Additionally, it can be observed that as the channel estimation error increases, the variance of channel dispersion also increases, meaning that imperfect CSI introduces more instability in channel performance. However, as the number of APs increases, the system with various channel estimation error approaches the deterministic channel, and the variance of channel dispersion tends to zero.

\subsection{Statistical Properties of Channel Capacity}\label{sec:3}
The expectation and variance of channel capacity with imperfect CSI can be derived in a manner similar to the process of solving for channel dispersion.

\textit{Theorem 3}: Accurate closed-form expressions for the expectation and variance of channel capacity based on Laplace approximation.

The expectation of channel capacity is expressed as:
\begin{equation}
\begin{aligned}
  {\mathbb E}\left[ {C( {\widehat{\bf{H}}} )} \right] &= \frac{M}{{\ln 2}}\int_0^\infty  {\frac{{{e^{ - s}}\left( {1 - {{\mathcal{L}}_X}\left( {{\beta ^{ - 1}}s} \right)} \right)}}{s}ds}.    \label{eq:LHSvarC}
\end{aligned}
\end{equation}

The variance of channel capacity is expressed as (\ref{eq:varCC}) at the top of the next page.
\begin{figure*}[ht]
\centering
\begin{equation}
\begin{aligned}
&\text{Var}\left[ {C( {\widehat{\bf{H}}} )} \right] = {\left( {\frac{M}{{\ln 2}}} \right)^2}\int_0^\infty  {\int_0^\infty  {\frac{{{e^{ - u - v}}\left( {{\mathcal{L}_X}\left( {{\beta ^{ - 1}}u + {\beta ^{ - 1}}v} \right) - {\mathcal{L}_X}\left( {{\beta ^{ - 1}}u} \right){\mathcal{L}_X}\left( {{\beta ^{ - 1}}v} \right)} \right)}}{{uv}}du} dv} \label{eq:varCC}
\end{aligned}
\end{equation}
\hrulefill
\end{figure*}

\textit{Proof}: See Appendix D.

 \begin{figure}[ht]
 \centering
 \includegraphics[width=3in]{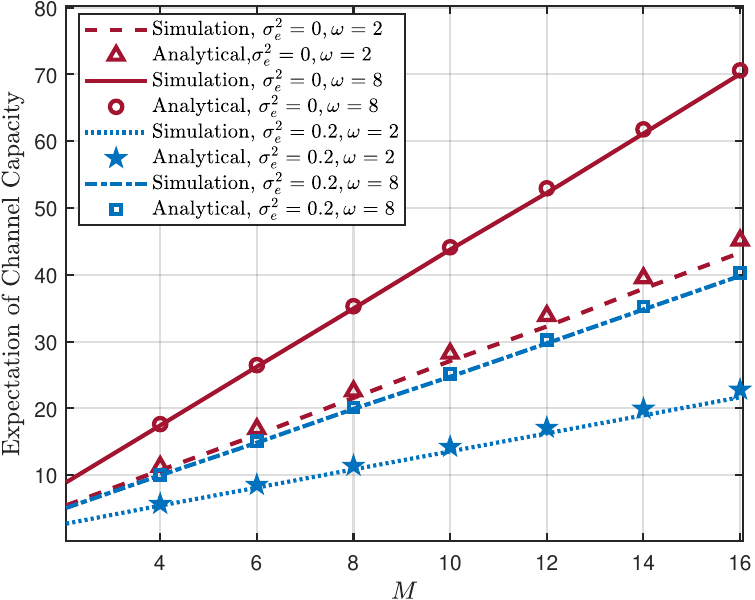}
 \caption{Performace comparison of simulation and analytical values of channel capacity with $E_N=8, \gamma=-20\ {\rm{dB}}, R=50m$}
 \label{fig3}
 \end{figure}

Fig. \ref{fig3} presents the simulation and analytical results for the expected channel capacity, with the analytical result calculated from (\ref{eq:LHSvarC}). Imperfect CSI can lead to a significant decrease in channel capacity. However, it can be observed that the reduction in channel capacity due to imperfect CSI can be mitigated by a higher $\omega$ or $M$, which represent two key advantages of mMIMO.

 \begin{figure}[t]
 \centering
 \includegraphics[width=3in]{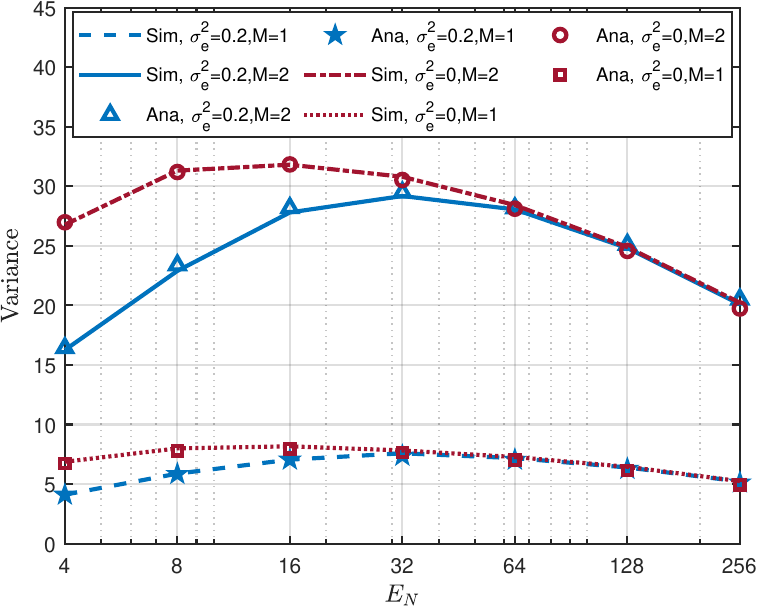}
 \caption{Performace comparison of simulation and analytical values of variance of channel capacity with $E_N=8, \gamma=-20 {\rm{dB}}, R=50 m$}
 \label{fig4}
 \end{figure}
Fig. \ref{fig4} presents the simulation and analytical results for the variance of channel capacity. It illustrates that the variance of channel capacity with imperfect CSI is lower than that with perfect CSI, due to the reduced channel capacity. As the spatial DoF increases, the variance of channel capacity increases as well. Furthermore, the variance of channel capacity tends to converge to the same value for both perfect and imperfect CSI, ultimately approaching zero for large $E_N$. This indicates that channels with various channel estimation errors converge to a deterministic channel as the number of APs increases, leveraging the advantages of cell-free architecture over centralized MIMO.

\section{System Performance Analysis Based on Closed-Form Approximations} \label{sec:4}

Building on the accurate expressions for the expected channel capacity (\ref{eq:LHSvarC}) and expected channel dispersion (\ref{eq:VN1}) derived in the previous sections, these equations reveal a clear relationship between imperfect CSI, $\omega$, and their collective impact on system performance, as they jointly influence the same parameter $\beta$ in expected channel capacity and expected channel dispersion. In this section, we delve into the impact of imperfect CSI on the system performanace and demonstrate that CF-mMIMO serves as a key architecture to mitigate the performace degradation caused by imperfect CSI.

\subsection{Average Achievable Rate}
The average achievable rate with imperfect CSI in finite blocklength regime can be expressed as:
\begin{equation}
{R^*}\left( {\tau ,\varepsilon } \right) \geqslant \overline R  = \mathbb{E}\left[ {C(\widehat {\mathbf{H}})} \right] - \sqrt {\frac{{\mathbb{E}\left[ {V(\widehat {\mathbf{H}})} \right]}}{\tau }} \frac{{{Q^{ - 1}}\left( \varepsilon  \right)}}{{\ln 2}},
\end{equation}
where  $\mathbb{E}\left[ {V(\widehat {\mathbf{H}})} \right]$ and $\mathbb{E}\left[ {C(\widehat {\mathbf{H}})} \right]$ are given by (\ref{eq:VN1}) and (\ref{eq:LHSvarC}), respectively. Thus, we have an accurate expression for the average achievable rate of CF-mMIMO with imperfect CSI in finite blocklength regime:
\begin{equation}
\begin{aligned}
\overline R  \approx  \frac{M}{{\ln 2}}\int_0^\infty & {\frac{{{e^{ - s}}\left( {1 - {\mathcal{L}_X}\left( {{\beta ^{ - 1}}s} \right)} \right)}}{s}ds}  \\
-& \sqrt {\frac{{M - M{\beta ^2}\int_0^\infty  s {e^{ - s\beta }}{\mathcal{L}_X}\left( s \right)ds}}{\tau }} \frac{{{Q^{ - 1}}\left( \varepsilon  \right)}}{{\ln 2}}. \label{eq:rateM}
\end{aligned}
\end{equation}

 \begin{figure}[t]
 \centering
 \includegraphics[width=3in]{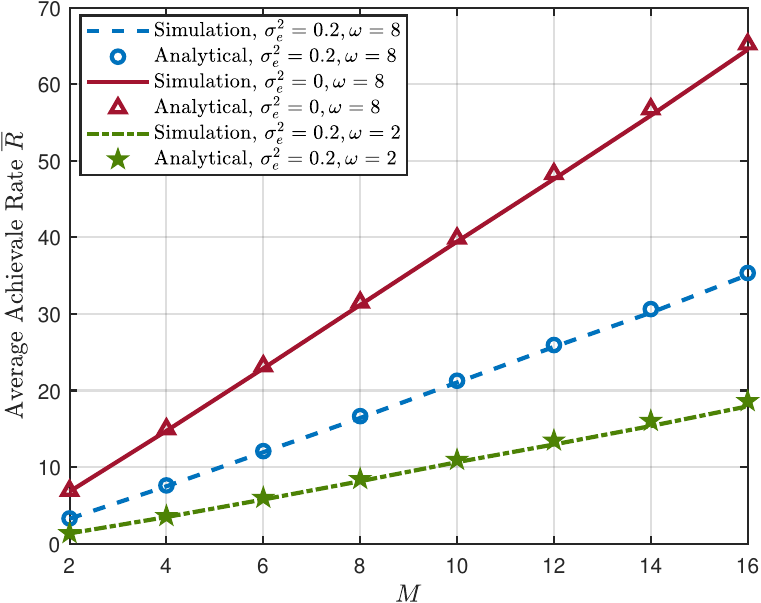}
 \caption{Performace comparison of simulation and analytical values of channel capacity with $E_N=8, \gamma=-20\ {\rm{dB}}, R=50 m,\tau=30, {\rm BEP}=10^{-7}$}
 \label{fig5}
 \end{figure}
Fig. \ref{fig5} shows the achievable rate for a given blocklength and BEP, demonstrating the performance benefits of increased spatial DoF in low-latency scenarios. It illustrates that imperfect CSI has a significant impact on the performance of the achievable rate. However, the reduction in the achievable rate due to the imperfect CSI can be compensated by a larger antenna array, demonstrating the advantages of mMIMO.

\subsection{Normalized Achievable Rate}
The channels between the typical user and different APs can be converted into multiple parallel links after performing eigenvalue decomposition. Therefore, the average achievable rate of each parallel link plays a crucial role in evaluating the performance of CF-mMIMO systems in finite blocklength regime. 

The normalized achievable rate can be obtained by dividing (\ref{eq:rateM}) by the spatial DoF:
\begin{equation}
\begin{aligned}
\frac{\overline R}{M}  \approx  \frac{1}{{\ln 2}}\int_0^\infty & {\frac{{{e^{ - s}}\left( {1 - {\mathcal{L}_X}\left( {{\beta ^{ - 1}}s} \right)} \right)}}{s}ds}  \\
-& \sqrt {\frac{{1 - {\beta ^2}\int_0^\infty  s {e^{ - s\beta }}{\mathcal{L}_X}\left( s \right)ds}}{M\tau }} \frac{{{Q^{ - 1}}\left( \varepsilon  \right)}}{{\ln 2}}. \label{eq:ratemM}
\end{aligned}
\end{equation}

 \begin{figure}[t]
 \centering
 \includegraphics[width=3 in]{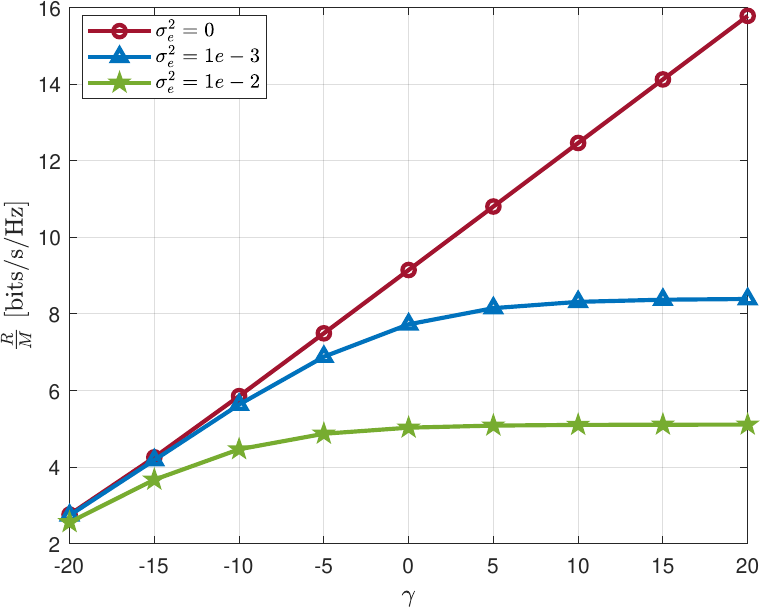}
 \caption{Comparison of normalized achievable rate with respect to SNR for various channel estimation errors with $E_N=8, R=50\ m,M=2,\omega=8,\tau=10, \text{BEP}= 10^{-7}$}
 \label{fig6}
 \end{figure}

Fig. \ref{fig6} shows the normalized achievable rate with various channel estimation errors for a given BEP and blocklength. It illustrates the continuous gain in the normalized achievable rate as a function of SNR with perfect CSI. However, when the CSI is imperfect, the growth of the normalized achievable rate becomes constrained, indicating that higher channel estimation errors negatively impact system performance even under high SNR conditions. It can be concluded that in scenarios with imperfect CSI, leveraging the advantages of CF-mMIMO architecture becomes a more effective approach to enhancing the normalized achievable rate.

 \begin{figure}[t]
 \centering
 \includegraphics[width=3 in]{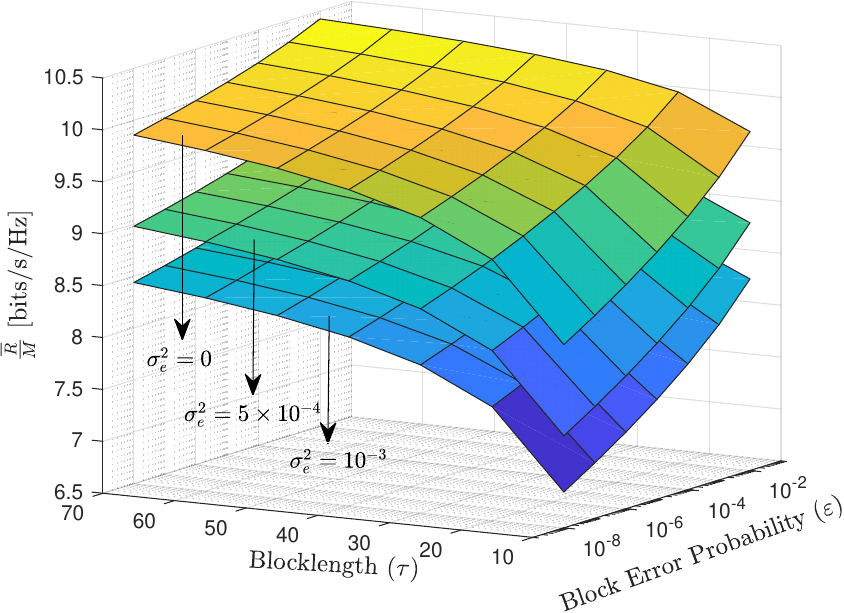}
 \caption{Comparison of normalized achievable rate for different channel estimation error with $E_N=8, \gamma=0 \ \rm{dB}, R=50\ m,M=1,\omega=8$}
 \label{fig7}
 \end{figure}

Furthermore, we explored the potential relationship between blocklength, BEP, and the normalized achievable rate under various CSI conditions in Fig. \ref{fig7}. It illustrates the normalized achievable rate as a function of $\tau$ and $\varepsilon$, with varying levels of channel estimation errors, enabling multiple surface comparisons. The results show that imperfect CSI significantly impacts the normalized achievable rate under the same system configuration parameters. Moreover, it is clear that imperfect CSI does not influence the channel capacity collapse effect \cite{6}.

\subsection{Block Error Probability}
\begin{equation}
\begin{aligned}
\varepsilon = Q\Bigg\{  \Bigg[ 
  \int_0^\infty \frac{e^{-s} \left( 1 - \mathcal{L}_X(\beta^{-1}s) \right)}{s} \, ds  - \frac{\bar{R}}{M}\ln 2 
  \Bigg] \nonumber \\
   \times \sqrt{\frac{M\tau}{1 - \beta^2 \int_0^\infty s e^{-s\beta} \mathcal{L}_X(s) \, ds}}
\Bigg\}
\end{aligned}
\end{equation}

 \begin{figure}[t]
 \centering
 \includegraphics[width=3 in]{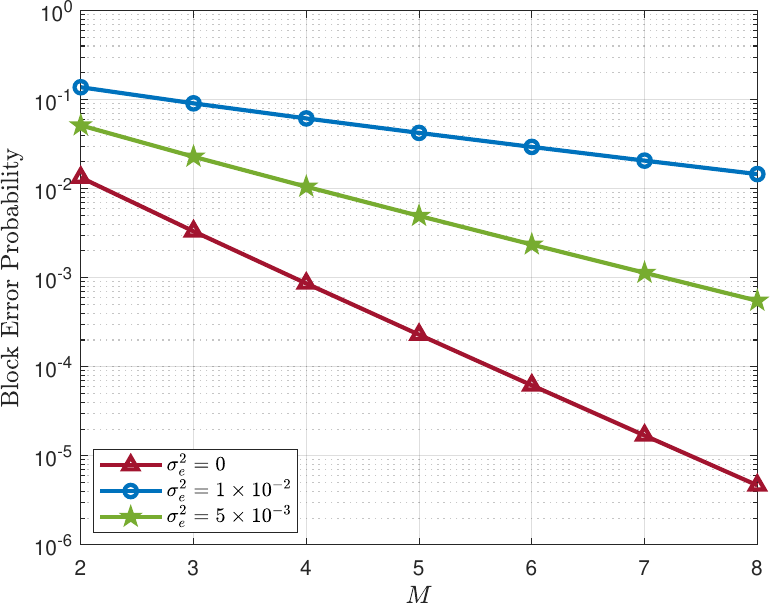}
 \caption{Comparison of block error probability for different channel estimation error with $E_N=8, \gamma=-20 \ \rm{dB}, R=50\ m,\omega=8, \tau=30,\overline R/M=4$ bit/s/Hz coding rate per antenna}
 \label{fig8}
 \end{figure}

Fig. \ref{fig8} shows the BEP with various channel estimation errors for a given blocklength and normalized achievable rate, demonstrating the beneficial effect of increasing spatial DoF for low-latency scenarios. It can be seen that the increase in spatial DoF reduces the BEP for the same system configuration, and the increase in BEP due to the imperfect CSI can be compensated by a higher spatial DoF. This trend also holds true when deploying more APs. Therefore, the mMIMO technology and cell-free architecture have significant advantages for scenarios with imperfect CSI and finite blocklength.

\section{Conclusion}\label{sec:5}
In this paper, we derive the accurate expressions of channel capacity and channel dispersion of the CF-mMIMO system with imperfect CSI by Laplace approximation, which coincide well with the simulation results. We find that increasing the spatial DoF and deploying more APs can significantly improve the system performance in the finite blocklength regime. Moreover, this paper theoretically evaluates the potential of CF-mMIMO to mitigate the impact of imperfect CSI on system performance. We find that the mMIMO architecture can compensate for the performance degradation caused by imperfect CSI. Additionally, the cell-free architecture ensures that channels with different levels of channel estimation errors converge towards the same stability as deterministic channels. This evaluation allows us to intuitively observe the advantages of the cell-free architecture, providing a basis for parameter evaluation in the design of 6G CF-mMIMO systems for uRLLC transmission.

\appendices
\section{Preliminary Results}
\textit{Lemma 1}: Laplace transform approximation when $\beta$ and X are positve to ensure convergence of the Laplace transform.
\begin{equation}
\ln \left\{ {1 + {\beta ^{ - 1}}X} \right\} = \int_0^\infty  {\frac{{{e^{ - s}}\left( {1 - {e^{ - s{\beta ^{ - 1}}X}}} \right)}}{s}ds} ,\label{eq:L1}
\end{equation}
\begin{equation}
{\left( {\beta  + X} \right)^{ - 1}} = \int_0^\infty  {{e^{ - s\left( {\beta  + X} \right)}}} ds. \label{eq:L3}
\end{equation}
Take the derivative of $\beta$ in (\ref{eq:L3}),we have
\begin{equation}{\left( {\beta  + X} \right)^{ - 2}} = \int_0^\infty  {s{e^{ - s\left( {\beta  + X} \right)}}} ds,\end{equation} \label{eq:L2}
\begin{equation}{\left( {\beta  + X} \right)^{ - 4}} = \frac{1}{6}\int_0^\infty  {{s^3}{e^{ - s\left( {\beta  + X} \right)}}} ds.\end{equation} \label{eq:L4}

\textit{Lemma 2} \cite{11}: For the small-scale matrix ${\bf h}_n \in \mathbb{C}^{L \times M}, n = 1,...,N$, ${ {\bf{h}}_n^{\rm H}{{ {\bf{h}}}_n}} \in \mathbb{C}^{M\times M}$ is a Wishart distribution with degrees of freedom $L$ and scale matrix $\bf \Sigma$. 

The expectation property of the Wishart distribution:

\begin{equation}
{{\mathbb E}}\left[ {\left( {\widehat {\bf{h}}_n^{\rm H}{{\widehat {\bf{h}}}_n}} \right)} \right] = L {\bf \Sigma} \label{eq:Wishart}.
\end{equation}

The expected value of the logarithm of the determinant of a Wishart matrix:
\begin{equation}
{{\mathbb E}}\left[ {\ln \det \left( {\widehat {\bf{h}}_n^{\rm H}{{\widehat {\bf{h}}}_n}} \right)} \right] = \sum\limits_{m = 1}^M {\psi \left( {L - m + 1} \right)}  + \ln \det {\bf \Sigma}. \label{eq:Wishart2}
\end{equation}
where, $\psi\left(m\right) = \psi\left(1\right) + \sum_{i=1}^{m-1}\frac{1}{i}$ is Euler's digamma function and $\psi\left(-1\right) \approx 0.577215$ is the Eular-Mascheroni constant.

\textit{Lemma 3}\cite{6}: For large $\omega = L/M$, where $L$ and $M$ are integers and $L > M$, we have the approximation : 
\begin{equation}
\exp \left[ {\frac{1}{M}\sum\limits_{m = 1}^M {\psi \left( {L - m + 1} \right)}}  \right] \approx L.
\end{equation}

\section{Proof of Theorem 1}
For a homogeneous poisson point processes $\Phi$ with density $\lambda$ in a circular region $\mathcal A$ of radius $R$, we can derive the Laplace transform of $X \triangleq \sum_{n \in \Phi } {l\left( {{d_n}} \right)}$ on the fact that $\Phi \left(\mathcal A\right)=N$ , these $N$ points are independently and uniformly located in $\mathcal A$:
\begin{equation}
\begin{aligned}
  {{\mathcal{L}}_X}\left( s \right) &= {\mathbb E}\left[ {{e^{ - sX}}} \right] \\
&= {\mathbb E}\left[ {{e^{ - s\sum_{n \in \Phi } {l\left( {{d_n}} \right)} }}} \right] \\
&= {\mathbb E}\left[ {\prod\limits_{n \in \Phi } {{e^{ - sl\left( {{d_n}} \right)}}} } \right] \\ 
   &\buildrel (a) \over = \exp \left\{ { - \lambda \int_0^R {\left[ {1 - {e^{ - sl\left( r \right)}}} \right]2\pi rdr} } \right\} \\
   & = \exp \left\{ { - \lambda \left[ {\pi \left( {{R^2} - {e^{ - s}}} \right) - 2\pi \int_1^R {{e^{ - s{r^{ - \alpha }}}}rdr} } \right]} \right\}, \label{eq:Lx}\\
\end{aligned}
\end{equation}
step $(a)$ holds based on theorem on probability generating functions of PPPs \cite{10}.

Note that the intergral item of (\ref{eq:Lx}) can be transformed into incompleted Gamma function $\Gamma\left(\alpha,x\right)=\int_x^\infty e^{-t}t^{\left(\alpha-1\right)}dt$, $\gamma\left(\alpha,x\right)=\int_0^x e^{-t}t^{\left(\alpha-1\right)}dt$, we have
\begin{small}
\begin{equation}
\begin{aligned}
  2\pi \int_1^R {{e^{ - s{r^{ - \alpha }}}}rdr} \buildrel(a) \over= & \frac{{2\pi }}{\alpha }{s^{\frac{2}{\alpha }}}\int_{s{R^{ - \alpha }}}^s {{e^{ - u}}{u^{ - \frac{2}{\alpha } - 1}}du}  \\
  \buildrel \over = &\frac{{2\pi }}{\alpha }{s^{\frac{2}{\alpha }}}\Gamma \left( { - \frac{2}{\alpha },s{R^{ - \alpha }}} \right) - \frac{{2\pi }}{\alpha }{s^{\frac{2}{\alpha }}}\Gamma \left( { - \frac{2}{\alpha },s} \right) \\
   \buildrel(b) \over\approx& \frac{{2\pi }}{\alpha }{s^{\frac{2}{\alpha }}}\Gamma \left( { - \frac{2}{\alpha },s{R^{ - \alpha }}} \right) \cr 
  \buildrel(c) \over = &\frac{{2\pi }}{\alpha }{s^{\frac{2}{\alpha }}}\Gamma \left( { - \frac{2}{\alpha }} \right) - \frac{{2\pi }}{\alpha }{s^{\frac{2}{\alpha }}}\gamma \left( { - \frac{2}{\alpha },s{R^{ - \alpha }}} \right),   \label{eq:Lx1}
\end{aligned}
\end{equation}
\end{small}
where step $(a)$ holds for variable substitution ${s{r^{ - \alpha }} = u} $. Step $(b)$ holds for the fact that when $R \gg 1$, we have $s{R^{ - \alpha }} \ll s$, thus $\Gamma \left( { - \frac{2}{\alpha },s} \right) \ll \Gamma \left( { - \frac{2}{\alpha },s{R^{ - \alpha }}} \right)$,
Step $(c)$ holds based on the complementary property of the incomplete Gamma function $\Gamma \left( {\alpha ,x} \right) + \gamma \left( {\alpha ,x} \right) = \Gamma \left( \alpha  \right)$.

Then, using the series expansion of the incomplete Gamma function, we can obtain
\begin{equation}
\begin{aligned}
{s^{\frac{2}{\alpha }}}\gamma \left( { - \frac{2}{\alpha },s{R^{ - \alpha }}} \right) =& {s^{\frac{2}{\alpha }}}\sum\limits_{n = 0}^\infty  {\frac{{{{\left( { - 1} \right)}^n}{{\left( {s{R^{ - \alpha }}} \right)}^{n - \frac{2}{\alpha }}}}}{{n!\left( {n - 2/\alpha } \right)}}}  \\
=& \sum\limits_{n = 0}^\infty  {\frac{{{{\left( { - 1} \right)}^n}{{\left( s \right)}^n}{R^{2 - \alpha n}}}}{{n!\left( {n - 2/\alpha } \right)}}}  \\
\approx&  - \frac{\alpha }{2}{R^2} - \frac{{{R^{2 - \alpha }}}}{{1 - 2/\alpha }}s .  \label{eq:Lx2}
\end{aligned}
\end{equation}

Bringing (\ref{eq:Lx1}) and (\ref{eq:Lx2}) into (\ref{eq:Lx}), we have a simplified form of the Laplace transformation of variable $X$, which can help simplify the closed-form expression in some cases
\begin{equation}
{{\mathcal{L}}_X^{approx.}}\left( s \right) = \exp \left\{ {\frac{{2\pi \lambda }}{\alpha }\Gamma \left( { - \frac{2}{\alpha }} \right){s^{\frac{2}{\alpha }}} + \frac{{2\pi \lambda {R^{2 - \alpha }}}}{{\alpha  - 2}}s} \right\}.  \label{eq:eqLx}
\end{equation}

\section{Proof of Theorem 2}
\subsection{Expectation of Channel Dispersion}

By substituting (\ref{eq:eqV2}) into (\ref{eq:V}), we derive the closed-form expression of channel dispersion in an integral form with high-precision approximation:
\begin{equation}
\begin{aligned}
 {\mathbb E}\left[{V({\widehat {\bf H}} ) }\right] = M - M \beta^2 \int_0^\infty s e^{-s\beta}\mathcal{L}_X\left(s\right)ds .
\end{aligned}
\end{equation}

\subsection{Variance of Channel Dispersion}
The variance of channel dispersion is expressed as:
\begin{equation}
\begin{aligned}
  &\text{Var}\left[ {V( {\widehat{\bf{H}}})} \right] = {\mathbb E}\left\{ {{{\left[ {V( {\widehat{\bf{H}}} ) - {\mathbb E}\left[ {V( {\widehat{\bf{H}}})} \right]} \right]}^2}} \right\}.  \label{eq:VNN}
\end{aligned}
\end{equation}

Substituting (\ref{eq:V}) into (\ref{eq:VNN}), we derive:
\begin{equation}
\begin{aligned}
   \text{Var}\left[ {V( {\widehat{\bf{H}}} )} \right] = & {\mathbb E}\left[ {{G^2}} \right] - \left( {\mathbb E} \left[ G \right] \right)^2.
\end{aligned}
\end{equation}

It is observed that solving for the first expectation, ${\mathbb E}\left[ {{G^2}} \right]$, is sufficient to determine the variance of the channel dispersion, as ${\mathbb E} \left[ G \right]$ is already included in the second term on the RHS of (\ref{eq:VN1}). Following the approach in (\ref{eq:eq001}), we have
\begin{equation}
\begin{aligned}
  &{{\mathbb E}_{\bf{h}}}\left[ {{G^2}} \right] \\
&= {{\mathbb E}_{\bf{h}}}\left\{ {{\text{t}}{{\text{r}}^2}\left[ {{{\left( {{{\bf{I}}_M} + \frac{\rho }{{M + M\rho \sigma _e^2\theta }}\sum\limits_{n \in \Phi } {l\left( {{d_n}} \right)\widehat {\bf{h}}_n^{\rm H}{{\widehat {\bf{h}}}_n}} } \right)}^{ - 2}}} \right]} \right\} \\
   &\buildrel \left(a\right) \over \approx {\text{t}}{{\text{r}}^2}\left[ {{{\left( {{{\bf{I}}_M} + \frac{L}{M}\frac{{\rho \left( {1 - \sigma _e^2} \right)}}{{1 + \rho \sigma _e^2\theta }}\sum\limits_{n \in \Phi } {l\left( {{d_n}} \right){{\bf{I}}_M}} } \right)}^{ - 2}}} \right] \\ 
   &= {M^2}{\left[ {1 + \frac{L}{M}\frac{\rho\left( {1 - \sigma _e^2} \right) }{{1 + \rho \sigma _e^2\theta }}\sum\limits_{n \in \Phi } {l\left( {{d_n}} \right)} } \right]^{ - 4}} \\
   &= {M^2}{\beta ^4}{\left[ {\beta  + \sum\limits_{n \in {\Phi}} {l\left( {{d_n}} \right)} } \right]^{ - 4}} , \label{eq:e333}
\end{aligned}
\end{equation}
where, step $\left(a\right)$ holds due to the Cauchy-Schwarz inequality. Based on Lemma 1, the Laplace approximation is applied to (\ref{eq:e333}) as follows:
\begin{equation}
\begin{aligned}
  {\mathbb E}\left[ {{G^2}} \right] &=\frac{{M^2}{\beta ^4}}{6}\int_0^\infty  {{s^3}{e^{ - s\beta }}{{\mathcal{L}}_X}\left( s \right)} ds. \label{eq:qq}
\end{aligned}
\end{equation}

By substituting (\ref{eq:eqV2}) and (\ref{eq:qq}) into (\ref{eq:VNN}), we can derive the closed-form expression for the variance of channel dispersion:
\begin{equation}
\begin{aligned}
{\text{Var}}\left[ {V( {\widehat {\bf{H}}} )} \right] = &\frac{{{M^2}{\beta ^4}}}{6}\int_0^\infty  {{s^3}{e^{ - s\beta }}{{\mathcal{L}}_X}\left( s \right)} ds\\
&- {M^2}{\beta ^4}{\left[ {\int_0^\infty  {s{e^{ - s\beta }}{{\mathcal{L}}_X}\left( s \right)} ds} \right]^2}.
\end{aligned}
\end{equation}

\subsection{Simplified Expectation of Channel Dispersion}
By using the simplified Laplace transform (\ref{eq:eqV4}), we have
\begin{equation}
\begin{aligned}
 {\mathbb E}\left[{V({\widehat {\bf H}} ) }\right] =M-M \beta^2 {\int_0^\infty  {s{e^{ - s\beta }}{{\mathcal{L}}_X^{approx.}}\left( s \right)} ds}
\label{eq:VSI}
\end{aligned}
\end{equation}
where, we define 
\begin{equation}
\begin{aligned}
&G1 \triangleq {\int_0^\infty  {s{e^{ - s\beta }}{{\mathcal{L}}_X^{approx.}}\left( s \right)} ds}\\
&=\int_0^\infty s \exp \left\{ \left( \frac{2 \pi \lambda R^{2 - \alpha}}{\alpha - 2} - \beta \right)s+ \frac{2 \pi \lambda}{\alpha} \Gamma \left( -\frac{2}{\alpha} \right) s^{\frac{2}{\alpha}} \right\}ds
\end{aligned}
\end{equation}

By analyzing the integral $G1$, we observe that the structure of the exponential and polynomial terms suggests a relationship with the Gamma function. To further eliminate the integral and derive a simplified closed-form expression, a Taylor expansion is applied to $G1$ as follows:
\begin{small}
\begin{equation}
\begin{aligned}
   &G1 \\
&=  \int_0^\infty s \exp \left\{ \left( \frac{2\pi \lambda R^{2-\alpha}}{\alpha - 2} - \beta \right) s \right\} \exp \left\{ \frac{2\pi \lambda}{\alpha} \Gamma \left( - \frac{2}{\alpha} \right) s^{\frac{2}{\alpha}} \right\} \, ds \\
  &\buildrel (a) \over = \int_0^\infty \sum_{k=0}^{\infty} \frac{\left( \frac{2\pi \lambda}{\alpha} \Gamma \left( - \frac{2}{\alpha} \right) \right)^k}{k!} s^{\frac{2k}{\alpha}+1} \exp \left\{ \left( \frac{2\pi \lambda R^{2-\alpha}}{\alpha - 2} - \beta \right) s \right\} \, ds \\
 & =  \sum_{k=0}^{\infty} \frac{\left( \frac{2\pi \lambda}{\alpha} \Gamma \left( - \frac{2}{\alpha} \right) \right)^k \Gamma \left( \frac{2k}{\alpha} + 2 \right)}{\left( \beta - \frac{2\pi \lambda R^{2-\alpha}}{\alpha - 2} \right)^{\frac{2k}{\alpha} + 2} k!}\\
 & \buildrel (b) \over =  \frac{1}{\left( \beta - \frac{2\pi \lambda R^{2-\alpha}}{\alpha - 2} \right)^2} {}_1\psi_0 \left[ \left( 2, \frac{2}{\alpha} \right); \frac{\frac{2\pi \lambda}{\alpha} \Gamma \left( - \frac{2}{\alpha} \right)}{\left( \beta - \frac{2\pi \lambda R^{2-\alpha}}{\alpha - 2} \right)^{\frac{2}{\alpha}}} \right]. \label{eq:G1}
\end{aligned}
\end{equation}
\end{small}
where, step $\left(a\right)$ utilizes the Taylor expansion of the exponential function, while step $\left(b\right)$ represents the summation of the Gamma function using the Fox-H function form represented as 
\begin{small}
${}_p\psi_q \left[ \begin{matrix} \left(a_1, A_1\right), \dots, \left(a_p, A_p\right) \\ \left(b_1, B_1\right), \dots, \left(b_q, B_q\right) \end{matrix} ; z \right]=\sum_{n=0}^{\infty} \frac{\Gamma(a_1 + A_1 n) \cdots \Gamma(a_p + A_p n)}{\Gamma(b_1 + B_1 n) \cdots \Gamma(b_q + B_q n)} \frac{z^n}{n!}.$
\end{small}

Thus a closed-form expression related to the expectation of channel dispersion with minimal loss of precision is derived as:
\begin{equation}
\begin{aligned}
 {\mathbb E}\left[{V({\widehat {\bf H}}) }\right] &= M - \frac{M \beta^2}{\left( \beta - \frac{2 \pi \lambda R^{2 - \alpha}}{\alpha - 2} \right)^2} \\
&\quad \times {}_1 \psi_0 \left[ \left( 2, \frac{2}{\alpha} \right); \frac{\frac{2 \pi \lambda}{\alpha} \Gamma\left( - \frac{2}{\alpha} \right)}{\left( \beta - \frac{2 \pi \lambda R^{2 - \alpha}}{\alpha - 2} \right)^{\frac{2}{\alpha}}} \right] 
\end{aligned}
\end{equation}

\subsection{Simplified Variance of Channel Dispersion}

By using the simplified Laplace transform (\ref{eq:eqV4}), we have
\begin{equation}
\begin{aligned}
{\text{Var}}\left[ {V( {\widehat {\bf{H}}} )} \right] = &\frac{{{M^2}{\beta ^4}}}{6}\int_0^\infty  {{s^3}{e^{ - s\beta }}{{\mathcal{L}}_X^{approx.}}\left( s \right)} ds\\
&- {M^2}{\beta ^4}{\left[ {\int_0^\infty  {s{e^{ - s\beta }}{{\mathcal{L}}_X^{approx.}}\left( s \right)} ds} \right]^2}. \label{eq:VVVG}
\end{aligned}
\end{equation}
where, we define 
\begin{equation}
\begin{aligned}
G2 &\triangleq \int_0^\infty  {{s^3}{e^{ - s\beta }}{{\mathcal{L}}_X^{approx.}}\left( s \right)} ds\\
&=\int_0^\infty s^3 \exp \left\{ \left( \frac{2 \pi \lambda R^{2 - \alpha}}{\alpha - 2} - \beta \right)s+ \frac{2 \pi \lambda}{\alpha} \Gamma \left( -\frac{2}{\alpha} \right) s^{\frac{2}{\alpha}} \right\}ds
\end{aligned}
\end{equation}

The expression form and solution method of G2 are similar to those for G1. The derivation details are omitted, and the result is:
\begin{equation}
\begin{aligned}
   G2 &= \frac{1}{ \left( \beta - \frac{2\pi \lambda R^{2 - \alpha}}{\alpha - 2} \right)^4} {}_1\psi_0 \left[  \left( 4, \frac{2}{\alpha} \right)  ; \frac{\frac{2\pi \lambda}{\alpha} \Gamma \left( -\frac{2}{\alpha} \right)}{\left( \beta - \frac{2\pi \lambda R^{2 - \alpha}}{\alpha - 2} \right)^{\frac{2}{\alpha}}} \right] \label{eq:G2}
\end{aligned}
\end{equation}

Thus, a simplified closed-form expression related to the variance of channel dispersion can be derived by substituting (\ref{eq:G2}) and (\ref{eq:G1}) into (\ref{eq:VVVG}).

\section{Proof of Theorem 3}
\subsection{Capacity of Channel Capacity}
As stated in (\ref{eq:C}), the expectation of channel capacity is
\begin{small}
\begin{equation}
\begin{aligned}
&C( {\widehat{\bf{H}}} ) \\
&= M{\log _2}{\left\{ {\det \left[ {{{\bf{I}}_M} + \sum\limits_{n \in {\Phi}} {\frac{\rho }{{M + M\rho \sigma _e^2\theta }}{{\widehat {\bf{H}}}_n}{{\left( {{d_n}} \right)}^{\rm H}}{{\widehat {\bf{H}}}_n}\left( {{d_n}} \right)} } \right]} \right\}^{\frac{1}{M}}} \\  
 &\buildrel (a) \over \geqslant M{\log _2}\left\{ {1 + \frac{\rho }{{M + M\rho \sigma _e^2\theta }}\sum\limits_{n \in {\Phi}} {{{\left[ {\det \left( {{{\widehat {\bf{H}}}_n}{{\left( {{d_n}} \right)}^{\rm H}}{{\widehat {\bf{H}}}_n}\left( {{d_n}} \right)} \right)} \right]}^{\frac{1}{M}}}} } \right\}, \label{eq:C2}
\end{aligned}
\end{equation}
\end{small}
where step $\left(a\right)$ applies Minkowski’s inequality \cite{28}. By substituting the expressions of large-scale and small-scale into (\ref{eq:C2}), the ergodic small-scale information can be expressed as
\begin{small}
\begin{equation}
\begin{aligned}
 & {{\mathbb E}_{\bf{h}}}\left[ {C({\widehat {\bf{H}}} )} \right] \\
   &\geqslant M{{\mathbb E}_{\bf{h}}}\left\{ {{{\log }_2}\left[ {1 + \frac{\rho }{{M + M\rho \sigma _e^2\theta }}\sum\limits_{n \in {\Phi}} {l\left( {{d_n}} \right)\exp \left[ {\frac{1}{M}\psi\left({{\widehat {\bf{h}}}_n}\right)} \right]} } \right]} \right\} \\
   &\buildrel (a) \over \geqslant M{\log _2}\left\{ {1 + \frac{\rho }{{M + M\rho \sigma _e^2\theta }}\sum\limits_{n \in {\Phi}} {l\left( {{d_n}} \right)\exp \left[ {\frac{1}{M}{{\mathbb E}_{\bf{h}}}\left\{ {\psi\left({{\widehat {\bf{h}}}_n}\right)} \right\}} \right]} } \right\}. \label{eq:V3}
\end{aligned}
\end{equation}
\end{small}
where, $\psi\left({{\widehat {\bf{h}}}_n}\right) \triangleq \ln \det ( {\widehat {\bf{h}}_n^{\rm H}{{\widehat {\bf{h}}}_n}} )$. Since ${\rm{ln}}\left(1+\sum_{n \in {\Phi}}{a_i}e^{x_i} \right)$ is a convex function of $x_i$ for $a_i \geqslant 0$. Using Jensen's inequality, step $\left(a\right)$ can be proved. 
Note that ${\widehat {\bf{h}}_n^{\rm H}{{\widehat {\bf{h}}}_n}}$ is a Wishart distribution with number of parameters $L$ and scale matrix $\left(1-\sigma_e^2\right){{\bf{I}}_M}$, the ergodic small-scale information can be obtained using Lemma 2,
\begin{equation}
\begin{aligned}
&{{\mathbb E}_{\bf{h}}}\left[ {C({\widehat {\bf{H}}} )} \right]\\
 &\geqslant M{\log _2}\left\{ {1 + \frac{{\rho \left( {1 - \sigma _e^2} \right)}}{{M + M\rho \sigma _e^2\theta }}\sum\limits_{n \in {\Phi }} {l\left( {{d_n}} \right) \Psi\left(L,M\right)} } \right\} \\ 
 &\buildrel (a) \over \approx M{\log _2}\left\{ {1 + \frac{L}{M}\frac{{\rho \left( {1 - \sigma _e^2} \right)}}{{1 + \rho \sigma _e^2\theta }}\sum\limits_{n \in {\Phi}} {l\left( {{d_n}} \right)} } \right\} \\
 &= M{\log _2}\left\{ {1 + {\beta ^{ - 1}}\sum\limits_{n \in {\Phi}} {l\left( {{d_n}} \right)} } \right\} ,
\end{aligned}
\end{equation}
where, $\Psi\left(L,M\right)\buildrel \Delta \over = \exp \left[ {\frac{1}{M}\sum_{m = 1}^M {\psi \left( {L - m + 1} \right)} } \right]$. Using Lemma 3, the approximation step $\left(a\right)$ holds. Then, according to Lemma 1, a further ergodic solution for the spatial randomness of APs can be expressed as follows:
\begin{equation}
\begin{aligned}
  {\mathbb E}\left[ {C( {\widehat{\bf{H}}} )} \right] &= \frac{M}{{\ln 2}}{{\mathbb E}_{\bf{d}}}\left[\ln \left( {1 + {\beta ^{ - 1}}X} \right) \right] \\
&= \frac{M}{{\ln 2}}\int_0^\infty  {\frac{{{e^{ - s}}\left( {1 - {{\mathcal{L}}_X}\left( {{\beta ^{ - 1}}s} \right)} \right)}}{s}ds}.    \label{eq:LHSvarC1}
\end{aligned}
\end{equation}

\subsection{Variance of Channel Capacity}
Similar to the variance of the channel dispersion, we have
\begin{equation}
\begin{aligned}
\text{Var}\left[ {C( {\widehat {\bf{H}}} )} \right] = {\mathbb E}\left[ {C{{( {\widehat {\bf{H}}} )}^2}} \right] - {\left( {{\mathbb E}\left[ {C( {\widehat {\bf{H}}} )} \right]} \right)^2}. \label{eq:varC}
\end{aligned}
\end{equation}

Substituting (\ref{eq:RHSvarC}) and (\ref{eq:LHSvarC1}) into (\ref{eq:varC}) ,we obtain the closed-form expression of the variance of the channel capacity, given by (\ref{eq:varCCQ}) on the top of this page, which involves a double integration of the Laplace transform of the interference term. 

\begin{figure*}[ht]
\centering
\begin{equation}
\begin{aligned}
  {\mathbb E}\left[ {C{{( {\widehat {\bf{H}}} )}^2}} \right] &= {\left( {\frac{M}{{\ln 2}}} \right)^2}{\mathbb E}\left\{ {{{\ln }^2}\left[ {1 + \frac{L}{M}\frac{{\rho \left( {1 - \sigma _e^2} \right)}}{{1 + \rho \sigma _e^2\theta }}\sum\limits_{n = 1}^N {l\left( {{d_n}} \right)} } \right]} \right\} \cr 
   &= {\left( {\frac{M}{{\ln 2}}} \right)^2}{\mathbb E}\left\{ {{{\ln }^2}\left[ {1 + {\beta ^{ - 1}}X} \right]} \right\} \cr 
   &= {\left( {\frac{M}{{\ln 2}}} \right)^2}{\mathbb E}\left\{ {{{\left[ {\int_0^\infty  {\frac{{{e^{ - s}}\left( {1 - {e^{ - s{\beta ^{ - 1}}X}}} \right)}}{s}ds} } \right]}^2}} \right\} \cr 
   &= {\left( {\frac{M}{{\ln 2}}} \right)^2}\int_0^\infty  {\int_0^\infty  {\frac{{{e^{ - u - v}}\left( {1 - {\mathcal{L}_X}\left( {{\beta ^{ - 1}}u} \right) - {\mathcal{L}_X}\left( {{\beta ^{ - 1}}v} \right) + {\mathcal{L}_X}\left( {{\beta ^{ - 1}}u + {\beta ^{ - 1}}v} \right)} \right)}}{{uv}}du} dv}.  \label{eq:RHSvarC}
\end{aligned}
\end{equation}
\end{figure*}

\begin{figure*}[ht]
\centering
\begin{equation}
\begin{aligned}
&\text{Var}\left[ {C( {\widehat{\bf{H}}} )} \right] = {\left( {\frac{M}{{\ln 2}}} \right)^2}\int_0^\infty  {\int_0^\infty  {\frac{{{e^{ - u - v}}\left( {{\mathcal{L}_X}\left( {{\beta ^{ - 1}}u + {\beta ^{ - 1}}v} \right) - {\mathcal{L}_X}\left( {{\beta ^{ - 1}}u} \right){\mathcal{L}_X}\left( {{\beta ^{ - 1}}v} \right)} \right)}}{{uv}}du} dv}. \label{eq:varCCQ}
\end{aligned}
\end{equation}
\hrulefill
\end{figure*}

\bibliographystyle{IEEEtran}
\bibliography{reference}

\begin{thebibliography}{10}
\providecommand{\url}[1]{#1}
\csname url@samestyle\endcsname
\providecommand{\newblock}{\relax}
\providecommand{\bibinfo}[2]{#2}
\providecommand{\BIBentrySTDinterwordspacing}{\spaceskip=0pt\relax}
\providecommand{\BIBentryALTinterwordstretchfactor}{4}
\providecommand{\BIBentryALTinterwordspacing}{\spaceskip=\fontdimen2\font plus
\BIBentryALTinterwordstretchfactor\fontdimen3\font minus
  \fontdimen4\font\relax}
\providecommand{\BIBforeignlanguage}[2]{{%
\expandafter\ifx\csname l@#1\endcsname\relax
\typeout{** WARNING: IEEEtran.bst: No hyphenation pattern has been}%
\typeout{** loaded for the language `#1'. Using the pattern for}%
\typeout{** the default language instead.}%
\else
\language=\csname l@#1\endcsname
\fi
#2}}
\providecommand{\BIBdecl}{\relax}
\BIBdecl

\bibitem{14}
H.~Q. Ngo, G.~Interdonato, E.~G. Larsson, G.~Caire, and J.~G. Andrews,
  ``{Ultradense Cell-Free Massive MIMO for 6G: Technical Overview and Open
  Questions},'' \emph{Proceedings of the IEEE}, vol. 112, no.~7, pp. 805--831,
  2024.

\bibitem{15}
E.~Björnson and L.~Sanguinetti, ``{Cell-Free versus Cellular Massive MIMO:
  What Processing is Needed for Cell-Free to Win?}'' in \emph{2019 IEEE 20th
  International Workshop on Signal Processing Advances in Wireless
  Communications (SPAWC)}, 2019, pp. 1--5.

\bibitem{29}
Y.~Zhang, Q.~Hu, M.~Peng, Y.~Liu, G.~Zhang, and T.~Jiang, ``{Interdependent
  Cell-Free and Cellular Networks: Thinking the Role of Cell-Free Architecture
  for 6G},'' \emph{IEEE Network}, vol.~38, no.~5, pp. 247--254, 2024.

\bibitem{16}
H.~Q. Ngo, A.~Ashikhmin, H.~Yang, E.~G. Larsson, and T.~L. Marzetta,
  ``{Cell-Free Massive MIMO Versus Small Cells},'' \emph{IEEE Transactions on
  Wireless Communications}, vol.~16, no.~3, pp. 1834--1850, 2017.

\bibitem{18}
A.~Lancho, G.~Durisi, and L.~Sanguinetti, ``{Cell-Free Massive MIMO for URLLC:
  A Finite-Blocklength Analysis},'' \emph{IEEE Transactions on Wireless
  Communications}, vol.~22, no.~12, pp. 8723--8735, 2023.

\bibitem{12}
J.~Östman, G.~Durisi, E.~G. Ström, M.~C. Coşkun, and G.~Liva, ``{Short
  Packets Over Block-Memoryless Fading Channels: Pilot-Assisted or Noncoherent
  Transmission?}'' \emph{IEEE Transactions on Communications}, vol.~67, no.~2,
  pp. 1521--1536, 2019.

\bibitem{8}
C.~Potter, K.~Kosbar, and A.~Panagos, ``{On Achievable Rates for MIMO Systems
  with Imperfect Channel State Information in the Finite Length Regime},''
  \emph{IEEE Transactions on Communications}, vol.~61, no.~7, pp. 2772--2781,
  2013.

\bibitem{1}
T.~Yoo and A.~Goldsmith, ``{Capacity of fading MIMO channels with channel
  estimation error},'' in \emph{2004 IEEE International Conference on
  Communications (IEEE Cat. No. 04CH37577)}, vol.~2.\hskip 1em plus 0.5em minus
  0.4em\relax IEEE, 2004, pp. 808--813.

\bibitem{22}
T.~Yoo, E.~Yoon, and A.~Goldsmith, ``{MIMO capacity with channel uncertainty:
  Does feedback help?}'' in \emph{IEEE Global Telecommunications Conference,
  2004. GLOBECOM'04.}, vol.~1.\hskip 1em plus 0.5em minus 0.4em\relax IEEE,
  2004, pp. 96--100.

\bibitem{17}
E.~Björnson and L.~Sanguinetti, ``{Making Cell-Free Massive MIMO Competitive
  With MMSE Processing and Centralized Implementation},'' \emph{IEEE
  Transactions on Wireless Communications}, vol.~19, no.~1, pp. 77--90, 2020.

\bibitem{23}
H.~Ren, C.~Pan, Y.~Deng, M.~Elkashlan, and A.~Nallanathan, ``{Joint Pilot and
  Payload Power Allocation for Massive-MIMO-Enabled URLLC IIoT Networks},''
  \emph{IEEE Journal on Selected Areas in Communications}, vol.~38, no.~5, pp.
  816--830, 2020.

\bibitem{24}
H.~Li, Y.~Wang, C.~Sun, and Z.~Wang, ``{User-Centric Cell-Free Massive MIMO for
  IoT in Highly Dynamic Environments},'' \emph{IEEE Internet of Things
  Journal}, vol.~11, no.~5, pp. 8658--8675, 2024.

\bibitem{19}
G.~Durisi, T.~Koch, and P.~Popovski, ``{Toward Massive, Ultrareliable, and
  Low-Latency Wireless Communication With Short Packets},'' \emph{Proceedings
  of the IEEE}, vol. 104, no.~9, pp. 1711--1726, 2016.

\bibitem{20}
Y.~Polyanskiy, H.~V. Poor, and S.~Verdu, ``{Channel Coding Rate in the Finite
  Blocklength Regime},'' \emph{IEEE Transactions on Information Theory},
  vol.~56, no.~5, pp. 2307--2359, 2010.

\bibitem{2}
A.~Papazafeiropoulos, P.~Kourtessis, M.~D. Renzo, S.~Chatzinotas, and J.~M.
  Senior, ``{Performance Analysis of Cell-Free Massive MIMO Systems: A
  Stochastic Geometry Approach},'' \emph{IEEE Transactions on Vehicular
  Technology}, vol.~69, no.~4, pp. 3523--3537, 2020.

\bibitem{3}
X.~You, B.~Sheng, Y.~Huang, W.~Xu, C.~Zhang, D.~Wang, P.~Zhu, and C.~Ji,
  ``{Closed-Form Approximation for Performance Bound of Finite Blocklength
  Massive MIMO Transmission},'' \emph{IEEE Transactions on Communications},
  vol.~71, no.~12, pp. 6939--6951, 2023.

\bibitem{6}
F.~Ye, X.~You, J.~Li, C.~Zhang, P.~Zhu, D.~Wang, and Y.~Huang,
  ``{Implementation of 6G $\text{TK} \upmu$ Extreme Connectivity via Cell-Free
  Massive MIMO System: A Theoretical Evaluation},'' \emph{IEEE Transactions on
  Wireless Communications}, vol.~23, no.~11, pp. 17\,744--17\,760, 2024.

\bibitem{4}
W.~Yang, G.~Durisi, T.~Koch, and Y.~Polyanskiy, ``{Quasi-Static
  Multiple-Antenna Fading Channels at Finite Blocklength},'' \emph{IEEE
  Transactions on Information Theory}, vol.~60, no.~7, pp. 4232--4265, 2014.

\bibitem{7}
E.~Björnson and L.~Sanguinetti, ``{Scalable Cell-Free Massive MIMO Systems},''
  \emph{IEEE Transactions on Communications}, vol.~68, no.~7, pp. 4247--4261,
  2020.

\bibitem{5}
E.~Abbe, S.-L. Huang, and E.~Telatar, ``Proof of the outage probability
  conjecture for miso channels,'' in \emph{2012 IEEE Information Theory
  Workshop}.\hskip 1em plus 0.5em minus 0.4em\relax IEEE, 2012, pp. 65--69.

\bibitem{9}
T.~Yoo and A.~Goldsmith, ``{Capacity and power allocation for fading MIMO
  channels with channel estimation error},'' \emph{IEEE Transactions on
  Information Theory}, vol.~52, no.~5, pp. 2203--2214, 2006.

\bibitem{26}
M.~Hasan, ``{Generalized Wielandt and Cauchy-Schwarz inequalities},'' in
  \emph{Proceedings of the 2004 American Control Conference}, vol.~3, 2004, pp.
  2142--2147.

\bibitem{25}
F.~C. Klebaner, \emph{Introduction to stochastic calculus with
  applications}.\hskip 1em plus 0.5em minus 0.4em\relax World Scientific
  Publishing Company, 2012.

\bibitem{11}
A.~M. Tulino, S.~Verd{\'u} \emph{et~al.}, ``{Random matrix theory and wireless
  communications},'' \emph{Foundations and Trends{\textregistered} in
  Communications and Information Theory}, vol.~1, no.~1, pp. 1--182, 2004.

\bibitem{10}
U.~Schilcher, S.~Toumpis, M.~Haenggi, A.~Crismani, G.~Brandner, and
  C.~Bettstetter, ``{Interference Functionals in Poisson Networks},''
  \emph{IEEE Transactions on Information Theory}, vol.~62, no.~1, pp. 370--383,
  2016.

\bibitem{28}
C.~R. Johnson and R.~A. Horn, \emph{{Matrix analysis}}.\hskip 1em plus 0.5em
  minus 0.4em\relax Cambridge university press Cambridge, 1985.

\end{thebibliography}

\balance

\end{document}